\documentclass[10pt,conference]{IEEEtran}
\usepackage[T1]{fontenc}
\usepackage{newtxtext,newtxmath}
\usepackage{array}
\usepackage{textcomp}
\usepackage[ruled,noend,linesnumbered]{algorithm2e}
\usepackage{algpseudocode}
\usepackage{stfloats}
\usepackage{url}
\usepackage{verbatim}
\usepackage{graphicx}
\usepackage{booktabs}
\usepackage{verbatim}

\usepackage{ifpdf}
 \ifpdf
 \else
 \fi

\usepackage{cite}

\makeatletter
\let\NAT@parse\undefined
\makeatother
\usepackage{hyperref}

\usepackage[capitalize, noabbrev, nameinlink]{cleveref}
\crefformat{figure}{#2Fig.\,#1#3}
\crefmultiformat{figure}{#2Figs.\,#1#3}{ #2and\,#1#3}{and\,#2#1#3}{, #2#1#3}
\crefformat{table}{#2Tab.\,#1#3}
\crefformat{algorithm}{#2Algo.\,#1#3}
\crefformat{equation}{#2Equ.\,#1#3}
\crefformat{section}{#2Sec.\,#1#3}
\crefmultiformat{section}{#2Secs.\,#1#3}{ #2and\,#1#3}{and\,#2#1#3}{, #2#1#3}
\crefformat{line}{#2Ln.\,#1#3}
\crefmultiformat{line}{#2Lns.\,#1#3}{ #2and\,#1#3}{and\,#2#1#3}{, #2#1#3}
\usepackage[all]{hypcap}
\def\BibTeX{{\rm B\kern-.05em{\sc i\kern-.025em b}\kern-.08em
    T\kern-.1667em\lower.7ex\hbox{E}\kern-.125emX}}
\usepackage{balance}

\usepackage[english]{babel}

\usepackage[letterpaper,top=2cm,bottom=2cm,left=3cm,right=3cm,marginparwidth=1.75cm]{geometry}

\usepackage[newfloat]{minted} 

\setminted{
  framesep=\fboxsep,
  frame=lines,
  breaklines=true,
  breakafter={:,},
  breakafter={_},
}
\AtBeginEnvironment{minted}{\setlength{\parskip}{-6pt}}
\setmintedinline{breaklines,breakafter={_:,},breakbefore={_jIG},bgcolor={}}

\title{
Enabling Memory-efficient Im2win Convolution with Multi-precision Support on GPU CUDA and Tensor Cores
}

\author{
\IEEEauthorblockN{
    Xiang Fu\IEEEauthorrefmark{1}
    Jixiang Ma\IEEEauthorrefmark{1}
    Xinpeng Zhang\IEEEauthorrefmark{1}
    Peng Zhao\IEEEauthorrefmark{2}
    Shuai Lu\IEEEauthorrefmark{1}
    Xu T. Liu\IEEEauthorrefmark{3}\IEEEauthorrefmark{4}%\orcidicon{0000-0003-3980-9803}%\thanks{This work is not related to the author's position at Amazon.}
    }
    \IEEEauthorblockA{
    \IEEEauthorrefmark{1}Nanchang Hangkong University, China
    \IEEEauthorrefmark{2}EEO Tech, China
    \IEEEauthorrefmark{3}University of Washington, USA
    \IEEEauthorrefmark{4}AWS, USA
    }
}

\begin{document}
\maketitle

\begin{abstract}
%Convolution is a principal computational bottleneck in deep neural networks, and its efficiency depends on tight integration between algorithms and GPU hardware. Existing GPU convolution methods face clear limitations: GEMM-based approaches incur large memory overhead, direct convolution suffers from poor cache utilization, FFT-based methods favor only large kernels, and Winograd methods reduce computation for small filters but raise memory usage and instability at larger sizes.
%Without these limitations, this work extends the im2win paradigm—a universal, memory-efficient convolution method that preserves contiguous memory access for all kernel sizes—to operate efficiently using full precision on CUDA cores and half precision on tensor cores. To harness tensor cores, we adapt and augment the im2win convolution with novel kernel designs and incorporate advanced optimization strategies, including zig-zag memory access and asynchronous data movement, enabling efficient exploitation of hardware-accelerated half-precision matrix multiply-accumulate (HMMA) operations.
Convolution is a principal computational bottleneck in deep neural networks, and its efficiency depends on tight integration between algorithms and GPU hardware. Existing GPU convolution methods suffer from large memory overhead, poor cache utilization, limited effectiveness across kernel sizes, or numerical instability.
This work extends the im2win paradigm—a universal, memory-efficient convolution method with contiguous memory access for all kernel sizes—to run efficiently in full precision on CUDA cores and half precision on tensor cores. By introducing new kernel designs and optimizations such as zig-zag memory access and asynchronous data movement, im2win efficiently exploits hardware-accelerated half-precision matrix multiply-accumulate operations.
Across twelve CNN benchmarks, im2win achieves up to 2.8$\times$ higher TFLOPS than its CUDA core implementation, 1.4$\times$ higher than cuDNN, and 6.4$\times$ higher than GEMM-based convolution with cuBLAS, while using as little as 53\% and 35\% of their memory, respectively. These results establish im2win as a unified, high-performance convolution framework for modern GPU architectures.
%Across twelve CNN benchmarks, the proposed im2win convolution algorithm leverages tensor cores to achieve 2.8$\times$ higher TFLOPS compared to the CUDA core implementation. When evaluated against state-of-the-art libraries, im2win attains 1.4$\times$ higher TFLOPS than cuDNN and 6.4$\times$ higher than GEMM-based convolution using cuBLAS, while uses 53\% and 35\% of the memory of these baselines, respectively. These findings demonstrate that im2win delivers significant improvements in both performance and memory efficiency, establishing it as a unified, high-performance convolution framework for modern GPU architectures.
\end{abstract}

\section{Introduction}
%Deep learning has become a foundation of artificial intelligence, with Convolutional Neural Networks (CNNs) widely used for their ability to learn spatial features in tasks such as computer vision. Convolution operations dominate CNN computation, accounting for 50\%-90\% of execution time\cite{shufflenet}. Efficient convolution on GPUs requires leveraging both CUDA cores—which provide full-precision parallel processing—and tensor cores, specialized units introduced in NVIDIA’s Volta architecture in 2017 to accelerate mixed-precision matrix operations. Optimizing convolution across these heterogeneous cores is essential to fully exploit GPU capabilities, reduce memory usage, and improve neural network performance.
Convolutional Neural Networks (CNNs), foundational to deep learning for spatial tasks like computer vision, rely heavily on convolution operations that account for 50–90\% of execution time\cite{shufflenet}. Efficient GPU convolution demands leveraging CUDA cores for full-precision processing and tensor cores—introduced in NVIDIA’s Volta architecture—for mixed-precision acceleration, to maximize performance and minimize memory usage. 

To enhance efficiency across diverse hardware, there are convolution algorithms like direct convolution which has zero memory overhead but poor cache utilization\cite{direct_conv_simd_ppopp_2023}, FFT (Fast Fourier Transform) convolution which has speedups for large kernels but latency on small filters, and Winograd convolution which has computation reduction for small fixed kernels but numerical instability at larger sizes. Im2col-based GEMM methods leverage libraries like cuBLAS but suffer high memory overhead from data transformation\cite{mec} and irregular matrices, yielding suboptimal performance\cite{ibxsmm}. GPU memory constraints drive processing strategies—from PyTorch's single-batch to cuDNN's mini-batch—and Implicit GEMM on tensor cores for efficient mixed-precision convolution without explicit intermediates.

Unlike prior methods limited by memory overhead (im2col), poor cache utilization (direct), latency on small kernels (FFT), or numerical instability (Winograd), the im2win (image to window) convolution paradigm addresses the high memory overhead and poor locality issues in im2col-based and direct convolutions\cite{im2win_hpec_2022, lu_im2win_2023}. It achieves this by transforming the input tensor into an im2win tensor that enables contiguous memory access and data reuse, significantly reducing memory consumption compared to im2col. 

Inspired by im2win, we extend it for GPU architectures with multi-precision support, implementing full-precision kernels on CUDA cores and half-precision kernels on tensor cores. To maximize hardware utilization on tensor cores, we apply additional kernel designs and optimization techniques such as index precomputation, zig-zag memory access and asynchronous data movement. 
%We also perform an ablation study to quantify the contribution of core optimization.
With thorough evaluation on twelve state-of-the-art CNN benchmarks, we compare our im2win convolution against PyTorch’s GEMM-based convolution using cuBLAS and multiple cuDNN convolution variants. We further conduct an ablation study to evaluate the performance impact of individual optimization techniques.

To summarize, this paper makes the following \textbf{main contributions} and is organized as follows:

\indent 1) We extend the im2win convolution paradigm to support multi-precision execution, implementing full-precision kernels for CUDA cores and half-precision kernels for tensor cores in~\cref{sec:im2win}. 

\indent 2) We augment the im2win convolution on CUDA and tensor cores, along with a suite of optimizations %following the guidance of the Roofline model 
in~\cref{optimization on gpu}. We analyze the effect of core optimizations through an ablation study in~\cref{ablation study}.

\indent 3) We conduct a comprehensive experimental evaluation comparing optimized im2win convolution, cuDNN algorithms and PyTorch’s im2col-based convolution implementations across multiple precisions and on both CUDA and tensor cores in~\cref{experiment on gpu}. \footnote{Code - \hyperlink{https://github.com/TensorConv/im2win/tree/IJCNN_2026}{https://github.com/TensorConv/im2win/tree/IJCNN\_2026}
}

\section{Related Work}
\label{sec:related_work}

%This section reviews direct, GEMM-based, implicit GEMM-based, FFT-based, and Winograd convolution methods, as well as other relevant GPU convolution works. Due to the large volume of related literature, we highlight only selected representative studies. % in this paper.

Convolution optimization strategies diverge significantly based on the underlying hardware architecture. On CPUs, implementations typically rely on either optimized direct convolution, employing micro-kernel reordering and cache-friendly layouts to enhance locality\cite{zhang2018high}, or explicit GEMM-based methods (im2col), which transform convolution into matrix multiplication\cite{chellapilla_high_2006} and may split large matrices to manage memory overhead\cite{p-im2col}. In contrast, on GPUs, to overcome the bandwidth bottlenecks of explicit data expansion, implicit GEMM strategies are widely adopted to compute indices on-the-fly\cite{evaluatcudnn_2019}; furthermore, recent advancements leverage specialized hardware like Tensor Cores, to accelerate these operations with arbitrary precision support\cite{qummar2025skepu}.

\begin{figure*}[!ht]
	\centering
    \includegraphics[width=5.5in]{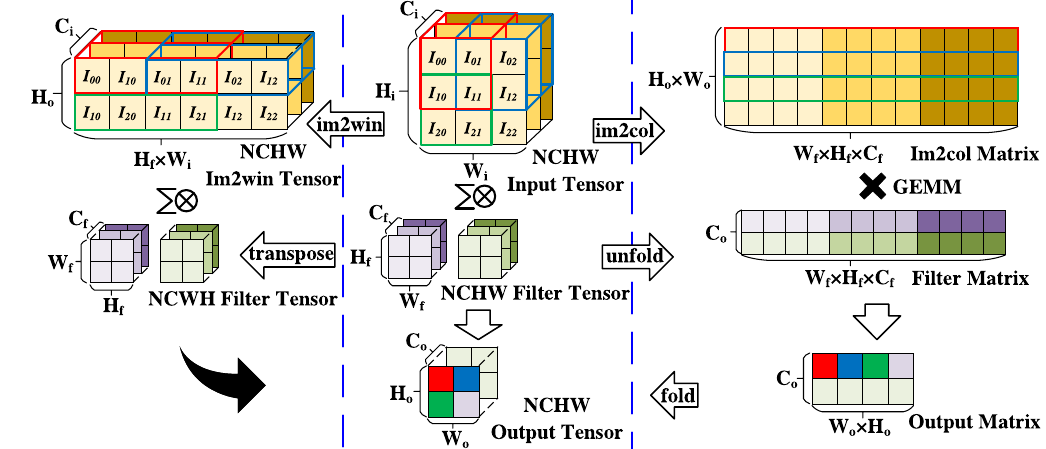}
	\caption{\small The schematic diagram illustrates, from left to right, direct convolution, im2win convolution, and im2col-based convolution, separated by blue dashed lines. From top to bottom, these represent the input (1×3×3×3×3), filter (2×3×2×2), and output (1×2×2×2). Colors denote distinct channels, while arrows indicate transformation processes. Red, blue, and green boxes within the input highlight the input regions required to compute the first three matching output elements. These regions are multiplied by the filter and summed to yield the corresponding output elements.}
	\label{fig:convolution}
\end{figure*}

\section{Im2win}
\label{sec:im2win}
To address the large memory usage of the im2col-based convolution and the inefficient, non-consecutive memory access of direct convolution, the im2win paradigm was proposed\cite{im2win_hpec_2022,lu_im2win_2023}. Im2win dramatically reduces memory overhead through a more compact data arrangement than im2col, enabling consecutive memory access and improved data reuse. This section describes the im2win paradigm, including its data transformation and convolution steps, and how we implement a high-performance im2win algorithm on tensor cores.
\subsection{Notation}
The convolution operation includes an input tensor ($\mathcal{I}$), a filter tensor ($\mathcal{F}$), and an output tensor ($\mathcal{O}$). In these tensors, $N$ is the batch size, $s$ is the stride size. Let $C_{i/f/o}$, $H_{i/f/o}$ and $W_{i/f/o}$ denote the channel, height and width of tensors respectively. The input, filter and output tensors in NCHW data layout are denoted as $\mathcal{I}[N_{i}][C_{i}][H_{i}][W_{i}]$, $\mathcal{F}[N_{f}][C_{f}][H_{f}][W_{f}]$ and $\mathcal{O}[N_{o}][C_{o}][H_{o}][W_{o}]$ respectively. The (2D) convolution is defined as:
\begin{equation}
\label{equation:convolution}
\begin{array}{r}
\begin{aligned}
\mathcal{O}_{(i, j, m, n)}=\sum_{r=1}^{C_f} \sum_{u=1}^{H_f} \sum_{v=1}^{W_f}\left(\mathcal{I}_{(i, r, m + u ,n + v)}\right. \\
\left.\times \mathcal{F}_{(j, r, u, v)}\right),
\end{aligned}
\end{array}
\end{equation}
where $1\leq i \leq N_o,1\leq j\leq C_o,1\leq m \leq H_o,1\leq n \leq W_o,1\leq r \leq C_f, 1\leq u \leq H_f, 1 \leq v \leq W_f$.

Tensors are often stored as one-dimensional arrays in memory, although logically represented as four-dimensional arrays, the cost of accessing different dimensions varies with different data layouts\cite{im2win_2024}. In the NCHW layout, the elements in the width dimension are contiguous in memory by prioritizing the width dimension first which means the cost of contiguous accessing the width dimension is the lowest, followed by height, channel and batch. NCHW and NHWC are more popular on CPUs\cite{dataLayout2}, while NCHW and CHWN are used on GPUs\cite{dataLayout1}.
 
\subsection{Im2win Data Transformation}
As shown in the red/blue/green boxes in~\cref{fig:convolution}, these input elements used to compute an output element are considered as a sliding window, The im2win transform enables the elements within these sliding windows to be arranged consecutively across different channels. The red boxes (in three channels) are the first window whose elements are used to compute the first element of the output tensor, the blue boxes the second, and the green boxes the third.  The im2win transformation is as follows:
\begin{equation}
 \mathcal{\hat{I}}(i,r,m,k \times H_f+u) = \mathcal{I}(i,r,m \times s + u,k),
\end{equation}
where $1\leq i \leq N_o,1\leq r\leq C_o,1\leq m \leq H_o,1\leq k \leq W_i,1\leq u \leq H_f$.

The upper right of~\cref{fig:convolution} illustrates the im2col transformation, where each row of the resulting matrix contains the elements of a sliding window. This approach leads to significant redundancy, as many elements are duplicated across neighboring rows, resulting in a much larger memory usage than the original input tensor. Compared to the im2col matrix, $\mathcal{\hat I}$ saves memory by eliminating redundant storage of overlapping elements. Specifically, it reduces the required space by $N_o \times C_o \times H_o \times W_o \times C_i \times H_f \times (W_f - s)$. Notably, when the stride $s$ equals the filter height $H_f$, the space used by the im2win tensor $\mathcal{\hat I}$ matches that of the im2col matrix, since there is no element reuse. However, as long as $s < H_f$, $\mathcal{\hat I}$ requires less memory than the im2col matrix. 
\subsection{Im2win Convolution}
A naive im2win convolution is illustrated in\cite{im2win_hpec_2022}, which uses a seven-layer nested loop structure like direct convolution. In~\cref{fig:convolution}, direct, im2win, and im2col convolutions are shown with their respective input formats at the top of the figure. For each output element, the algorithm multiplies the elements in a window (red box) with the corresponding filter elements and sums the results. The second output element takes the elements within the blue box as its input, representing the subsequent window.
In im2win convolution, the window shifts by $s \times H_f$ horizontally; for advancing to the next column of the output, the window moves down by 1. 
For each output channel $C_o$, the corresponding filter is selected.

Both direct and im2win convolutions compute all output elements by iterating through the outer four loops, with the innermost loops handling the detailed element-wise multiplications.
The operation can be formally described as:
\begin{equation}
\begin{array}{r}
\begin{aligned}
\mathcal{\hat I}_{(i, r, m, (n \times s) \times W_{f} + v \times W_{f} + u)}\times \mathcal{F}_{(j, r, u, v)},
\end{aligned}
\end{array}
\end{equation}
where the indices correspond to batch, channel, spatial, and filter dimensions.

The time complexity of im2win convolution is $O(N_o \times C_o \times H_o \times W_o \times C_f \times H_f \times W_f)$, which matches that of direct convolution. 
For filter data continuity, im2win leverages a filter tensor layout (such as NCHW) that prioritizes elements along the height before the width, aligning with the im2win access pattern.
Im2win further optimizes performance by reusing elements between overlapping windows in the same column, allowing elements loaded in a previous window to remain in cache for subsequent computations. This reduces memory loading time and improves overall efficiency compared to both direct and im2col-based convolutions.

\begin{algorithm}[t]
\small
\SetAlgoLined
\DontPrintSemicolon
\caption{\small High-Performance Im2win Convolution on Tensor Cores}
\label{algorithm:High_performance_im2win_tensor_core}
\KwIn{im2win input tensor $\hat{I}$, filter tensor $\mathcal{F}$, stride $s$}
\KwOut{output tensor $\mathcal{O}$}

\textbf{Dimensions}: 
$\mathbf{M}=C_o$, $\mathbf{N}=N_o \times H_o \times W_o$, $\mathbf{K} = C_f \times H_f \times W_f$ \;
\textbf{\# of blocks}: $M/M_B \times N/N_B$ \;
\label{algorithm:High_performance_im2win_tensor_core:line:block}
\textbf{\# number of threads }: $T$ \;
\label{algorithm:High_performance_im2win_tensor_core:line:threads}
\textbf{\# warp per block}: $M_B/M_W \times N_B/N_W$ \;
\label{algorithm:High_performance_im2win_tensor_core:line:warp}

\textbf{WMMA Registers}: $A_{\mathit{reg}}[2][N_W], B_{\mathit{reg}}[2][M_W], C_{\mathit{reg}}[M_W][N_W]$ \;
\label{algorithm:High_performance_im2win_tensor_core:line:dBuff}

\textbf{Shared memory}: $A_{\mathit{shm}}[2][K_W \times N_B], B_{\mathit{shm}}[2][K_W \times M_B]$ \;
\label{algorithm:High_performance_im2win_tensor_core:line:dBuff2}

$bx \gets (blockIdx.z \bmod 2 = 0)\ ?\ blockIdx.x\ :\ (gridDim.x - blockIdx.x)$\;
\label{algorithm:High_performance_im2win_tensor_core:line:zig}

$by \gets blockIdx.y + blockIdx.z \times gridDim.y$\;
\label{algorithm:High_performance_im2win_tensor_core:line:zig2}

$A_{\mathit{shm}}[0][K_W \times N_B / T]\underleftarrow{async\_vec\_load} (\mathcal{\hat{I}}(by, 0))$\;
\label{algorithm:High_performance_im2win_tensor_core:line:async_copy}
$B_{\mathit{shm}}[0][K_W \times M_B / T] \underleftarrow{async\_vec\_load}(\mathcal{F}(bx, 0))$\;
\label{algorithm:High_performance_im2win_tensor_core:line:async_copy2}
\textbf{commit\_and\_wait()}\\
\label{algorithm:High_performance_im2win_tensor_core:line:async_copy3}

wmma::load\_matrix\_sync($A_{\mathit{reg}}[0]$,$A_{\mathit{shm}}[0][K_W \times N_B]$)\;
\label{algorithm:High_performance_im2win_tensor_core:line:wmma_load}
    
wmma::load\_matrix\_sync($B_{\mathit{reg}}[0]$,$B_{\mathit{shm}}[0][K_W \times M_B]$)\;
\label{algorithm:High_performance_im2win_tensor_core:line:wmma_load2}

\For{$k  = 1$ \KwTo $K/K_W$ }{
    wmma::mma\_sync($C_{\mathit{reg}}$, $B_{\mathit{reg}}[write]$, $A_{\mathit{reg}}[write]$, $C_{\mathit{reg}}$) \;
    \label{algorithm:High_performance_im2win_tensor_core:line:wmma}

    \If{$k \neq K /K_W$}{
        $A_{\mathit{shm}}[read][K_W \times N_B / T] \underleftarrow{async\_vec\_load} (\mathcal{\hat{I}}(by, k))$\;\label{algorithm:High_performance_im2win_tensor_core:line:pre-fetch}

        $B_{\mathit{shm}}[read][K_W \times M_B / T] \underleftarrow{async\_vec\_load}(\mathcal{F}(bx, k))$\;\label{algorithm:High_performance_im2win_tensor_core:line:pre-fetch2}
        \textbf{commit\_and\_wait()}\\

        wmma::load\_matrix\_sync($A_{\mathit{reg}}[read]$ , $A_{\mathit{shm}}[read][K_W \times N_B]$) \;
    
        wmma::load\_matrix\_sync($B_{\mathit{reg}}[read]$ , $B_{\mathit{shm}}[read][K_W \times M_B]$) \;
    }
    \textbf{\_\_syncthreads()}\\
}

 $\mathcal{O}(bx, by)\underleftarrow{write}(C_{\mathit{reg}})$ \;

\end{algorithm}

\subsection{Im2win Convolution on Tensor Cores}
\label{subsec:im2win_tensor_core}
We implement the im2win convolution on tensor cores using the NHWC layout to enhance memory access contiguity for convolutional windows~\cite{im2win_2024}. The design aligns with the architectural features of CUDA and tensor cores, with careful consideration of thread-block organization, shared memory allocation, and warp-level computation. Shared memory is managed at the thread-block level, where each block contains multiple warps and threads. Data loading occurs at the thread level, while matrix multiplications on tensor cores are executed at the warp level.

We proposed high-performance im2win convolution algorithm implemented on tensor cores is presented in~\cref{algorithm:High_performance_im2win_tensor_core}. The algorithm begins by tiling the output tensor. Since the im2win convolution requires low extra memory, the im2win convolution can process all batches simultaneously to achieve higher parallelism. We map the output indices into two logical dimensions: $\mathbf{M}=C_o$ and $\mathbf{N}=N_o \times H_o \times W_o$. Each thread block is assigned a micro-tile of size $M_B\times N_B$ . As we use tensor cores via warp matrix-multiply accumulate (WMMA) instructions with 16$\times$16 granularity, both $M_B$ and $N_B$ are constrained to be multiples of 16. While larger blocks improve tensor cores utilization, they reduce the number of concurrent thread blocks and increase shared memory pressure. To strike a balance, we adopt two block shapes: 32$\times$128 and 64$\times$128.

For dimension $N$, we further decompose the mapping into two dimensions and traverse dimension $N$ in a zig-zag pattern by alternating parity order, thereby reducing memory bank conflicts and irregular access strides shown in~\cref{algorithm:High_performance_im2win_tensor_core:line:zig}. Within each thread block, computations are further distributed across multiple warps, with each warp responsible for an $M_W\times N_W$ tile. This ensures sufficient computation per warp to fully utilize tensor cores while maintaining adequate parallelism across warps.

Data in convolutional windows are loaded in a tiled manner. Input data required for convolution are fetched from global memory into shared memory using vectorized 16-byte load operations, with asynchronous data movement implemented through Parallel Thread Execution (PTX) instructions in~\cref{algorithm:High_performance_im2win_tensor_core:line:async_copy,algorithm:High_performance_im2win_tensor_core:line:async_copy2}. The data are moved from shared memory to registers using vectorized load operations in~\cref{algorithm:High_performance_im2win_tensor_core:line:wmma_load,algorithm:High_performance_im2win_tensor_core:line:wmma_load2}. 
To hide the data movement latency, we allocate two shared memory buffers and two registers in~\cref{algorithm:High_performance_im2win_tensor_core:line:dBuff,algorithm:High_performance_im2win_tensor_core:line:dBuff2}: one contains the data in current computation while the other pre-fetches for the next computation in~\cref{algorithm:High_performance_im2win_tensor_core:line:pre-fetch,algorithm:High_performance_im2win_tensor_core:line:pre-fetch2}. Upon computation, the results are written back to the output tensor.

\subsection{Optimizations for im2win convolutions on CUDA and Tensor Cores}
\label{optimization on gpu}
%This section details the optimization of im2win convolution, guided by the Roofline model\cite{samuel_roofline_2009} to maximize both computational throughput and memory bandwidth in line with tensor cores. 
%We also discuss about how to support multi-precision in im2win at the end of this section.
This subsection outlines optimizations for im2win convolutions, designed to fully harness the potential of tensor cores while maximizing computational throughput and memory bandwidth. We employed a suite of common optimization techniques, including: tiling and data pre-fetching, shared memory and register usage, vectorized load/store operations, double buffering. We further leveraged the capabilities of the NVIDIA SM 80 architecture by introducing two additional methods: asynchronous data movement and Zig-Zag access.
Owing to space constraints, common optimization techniques shall not be elaborated upon here; their specific implementation details may be found in the relevant literature \cite{lu_im2win_2023}.
\subsubsection{index precomputation}
As the index offsets required for each thread in the convolution are fixed, these offsets can be precomputed on the host and stored in GPU constant memory. This approach is primarily applied during the process of loading input tensors from global memory into shared memory, thereby avoiding redundant index calculations on the device and effectively reducing runtime overhead.
\subsubsection{Asynchronous data movement}
The main purpose of asynchronous data movement is to maximize computation utilization by separating the two tasks, data movement and data computation, so that they can be performed concurrently, thus hiding the data movement latency. We implement the underlying asynchronous data movement by using the PTX instruction. To further increase the utilization of the compute units and completely hide the data movement latency, we usually combine the asynchronous data movement with double buffering to build an efficient producer-consumer pipeline.
\subsubsection{Zig-Zag access}
In convolutional computation, when the thread block division of a certain dimension is too large, it easily leads to bank conflicts during shared memory access, which reduces the memory access efficiency. To alleviate this problem, the Zig-Zag access pattern can be used to rearrange the data access order\cite{volkov2008benchmarking}. We introduce an additional access dimension into the thread block division, and decide the access direction according to the parity of this dimension. This alternate access strategy significantly reduces the bank conflict caused by multiple thread blocks accessing the same address at the same time, and improves the throughput of parallel access.
\subsection{Multi-precision Support}
Convolution traditionally relies on single precision (FP32), half precision (FP16) offers superior memory bandwidth. and deliver higher GFLOPS rates \cite{gpnpu}. Our implementation supports two optimized variants: an FP32 version on CUDA cores and an FP16 version on tensor cores. On Ampere-based GPUs—the platform for our experiments—performing FP16 arithmetic without tensor cores often requires internal casting to FP32, which reduces performance benefits. Since Tensor Cores natively support efficient FP16 matrix multiply-accumulate (MMA), our FP16 variant leverages tensor cores using the WMMA API to achieve higher throughput.
\section{Experiments}
\label{experiment on gpu}

This section outlines the experimental setup, detailing the hardware configuration, benchmarks, and algorithms used for comparative analysis. Subsequently, it presents a comprehensive analysis of the GPU-based experimental results, focusing on key performance metrics such as TFLOPS and MaxRSS for each algorithm.

\subsection{Experimental Setup} 
\label{GPU setup}

\subsubsection{Experiment environment}
All experiments were conducted on a system equipped with an NVIDIA® GeForce RTX® 3090 GPU (24 GB memory, Ampere architecture featuring third-generation tensor cores), hosted on an Intel® Xeon® Silver 4214 CPU server. The implementation was built with CUDA 11.3 and integrated with PyTorch 2.2.0. For baseline comparisons, we evaluated against PyTorch’s GEMM-based convolution backend (linked to cuBLAS 11.3) as well as the convolution primitives in cuDNN 8.2.1. Peak GPU memory consumption was monitored in real time using the NVIDIA System Management Interface (nvidia-smi).
\subsubsection{cuDNN convolutions}
On CUDA cores, we study seven convolution implementations in cuDNN, which are grouped into im2col-based (im2col), implicit GEMM-based (IPG), FFT (FT), and Winograd (WN) approaches. 
Due to space limit, we present the results of the optimized variants of IPG, FT, and WN, omitting their less optimized ones.
cuDNN dynamically selects the most efficient algorithm at runtime based on hardware characteristics, stride, input and filter tensor properties. 
Consequently, the selected algorithm may vary across different runs. When leveraging tensor cores, cuDNN utilizes only the IPG and WN implementations.
\subsubsection{im2win convolutions}
We develop two variants of the im2win convolution: an FP32 implementation on CUDA cores and an FP16 one on tensor cores. The CUDA-core version builds upon prior work\cite{lu_im2win_2023}, enhanced with two new optimizations: index precomputation and vectorized data transfers between global and shared memory. 
We conduct an ablation study (see~\cref{ablation study}) with four implementations on tensor cores to evaluate the performance impact of individual one: a version incorporates all optimizations except index precomputation outlined in~\cref{optimization on gpu}, one without Zig-Zag access (wo\_zigzag), one without asynchronous data movement (wo\_async), and one without double buffering (wo\_double).
%However, not all optimizations consistently improve performance in the tensor cores setting; hence, 
%We conduct a detailed ablation study , which evaluates the performance impact of individual one.

\subsubsection{im2col convolutions with cuBLAS}
We compare with PyTorch’s implementation, backed by cuBLAS for the GEMM computation. We evaluate both FP32 on CUDA cores and FP16 on tensor cores. 

In all figures and tables in this section, “CC” denotes results obtained using FP32 precision executed on CUDA cores, while “TC” denotes results obtained using FP16 precision executed on tensor cores. 

\begin{table}[htbp]
\caption{\small Twelve convolutional layers of the DNN benchmarks}
\label{table:benchmarks}
\centering
\resizebox{\linewidth}{!}{
\begin{tabular}{@{}lcccccc@{}}
\toprule
\textbf{NAME}&\textbf{INPUT}&\textbf{FILTER, STRIDE}&\textbf{OUTPUT}\\
& $C_{i} \times H_{i} \times W_{i}$ & $C_{o} \times H_{f} \times W_{f}, s_{h}(s_{w})$&$C_{o} \times H_{o} \times W_{o}$\\
\midrule
$\textbf{cv1}$ & $3\times227\times227$& $96\times11\times11, 4$&$96\times55\times55$ \\
$\textbf{cv2}$ & $3\times231\times231$& $96\times11\times11, 4$&$96\times56\times56$ \\
$\textbf{cv3}$ & $3\times227\times227$& $64\times7\times7, 2$&$64\times111\times111$ \\
$\textbf{cv4}$ & $64\times224\times224$& $64\times7\times7, 2$&$64\times109\times109$ \\
$\textbf{cv5}$ & $96\times24\times24$& $256\times5\times5, 1$&$256\times20\times20$ \\
$\textbf{cv6}$ & $256\times12\times12$& $512\times3\times3, 1$&$512\times10\times10$ \\
$\textbf{cv7}$ & $3\times224\times224$& $64\times3\times3, 1$&$64\times222\times222$ \\
$\textbf{cv8}$ & $64\times112\times112$& $128\times3\times3, 1$&$128\times110\times110$ \\
$\textbf{cv9}$ & $64\times56\times56$& $64\times3\times3, 1$&$64\times54\times54$ \\
$\textbf{cv10}$ & $128\times28\times28$& $128\times3\times3, 1$&$128\times26\times26$ \\
$\textbf{cv11}$ & $256\times14\times14$& $256\times3\times3, 1$&$256\times12\times12$ \\
$\textbf{cv12}$ & $512\times7\times7$& $512\times3\times3, 1$&$512\times5\times5$ \\
\bottomrule
\end{tabular}}
\end{table}

\subsection{Benchmarks} 
To adequately cover the diversity of convolutional layers present in modern deep neural networks (DNNs), evaluating a single model with uniform filter dimensions—such as VGG-16, which exclusively employs 3$\times$3 kernels, or ResNet-50, which utilizes only three distinct kernel sizes—is insufficient. To address this limitation, our evaluation adopts a state-of-the-art DNN benchmark that includes twelve unique convolutional benchmarks (cv1–cv12),  detailed in~\cref{table:benchmarks}. 

\begin{table}[htbp]
\caption{\small the observed best performant cuDNN convolutions and im2win in each benchmark.}
\label{table:cuDNN_algo}
\centering
\resizebox{\linewidth}{!}{
\begin{tabular}{@{}lccc@{}}
\toprule
\textbf{NAME}&\textbf{CUDNN ALGO} & \textbf{CUDNN ALGO} & \textbf{IM2WIN ALGO}\\
&  \textbf{CUDA CORES} & \textbf{TENSOR CORES} & \textbf{TENSOR CORES}\\
\midrule
$\textbf{cv1}$  & IPG\_CC & IPG\_TC &baseline\\
$\textbf{cv2}$  & IPG\_CC & IPG\_TC &wo\_zigzag\\
$\textbf{cv3}$  & IPG\_CC & IPG\_TC &baseline\\
$\textbf{cv4}$ & FT\_CC & IPG\_TC &wo\_async\\
$\textbf{cv5}$ & WN\_CC & IPG\_TC &baseline\\
$\textbf{cv6}$ & WN\_CC & WN\_TC &baseline\\
$\textbf{cv7}$ & IPG\_CC & IPG\_TC &wo\_double\\
$\textbf{cv8}$ & FT\_CC & WN\_TC &baseline\\
$\textbf{cv9}$ & FT\_CC & IPG\_TC &baseline\\
$\textbf{cv10}$ & FT\_CC & WN\_TC &baseline\\
$\textbf{cv11}$ & WN\_CC & WN\_TC &wo\_async\\
$\textbf{cv12}$ & IPG\_CC & WN\_TC &baseline\\
\bottomrule
\end{tabular}}
\end{table}

\subsection{Overall Performance on CUDA and Tensor Cores}
\label{subsection:Overall Performance on GPU}

\begin{figure*}[!ht]
\centering
\includegraphics[width=\linewidth]{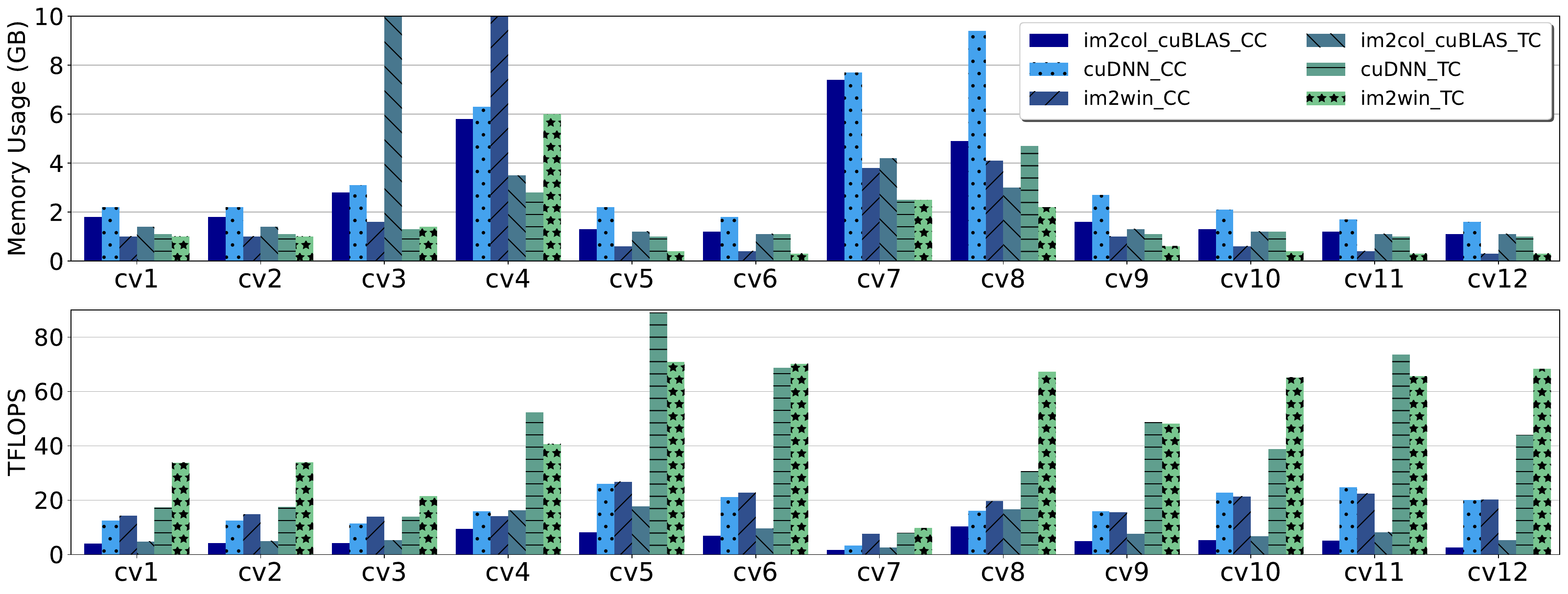}
\caption{\small Performance results in TFLOPS and memory usage among the three convolutional algorithms (im2win, im2col\_cuBLAS, and cuDNN) on CUDA and tensor cores respectively. Note: that the memory usage of im2col\_cuBLAS\_TC is 14.6 GB on cv3, and that of im2win\_CC is 11.7 GB on cv4.}
\label{fig:overall_performance}
\end{figure*}

We compare our im2win convolution with PyTorch’s im2col-based convolution with cuBLAS and the cuDNN convolutions. 
We set $N_i$=256 and run in each of the twelve convolutional benchmarks 50 times, reporting the highest observed TFLOPS and lowest MaxRSS.
Among four im2win\_TC versions, we report the version with the best performance for each benchmark, shown as the last column in \cref{table:cuDNN_algo}. Similarly, the auto-selected algorithm in cuDNN on CUDA and tensor cores are reported in the second and third columns respectively in \cref{table:cuDNN_algo}.

\cref{fig:overall_performance} summarizes the TFLOPS and memory usage of each method. 
Im2win\_TC achieves a 2.7$\times$ increase in overall TFLOPS compared to im2win\_CC. This performance gain arises from two main factors. First, im2win\_TC employs the FP16 data type, whereas im2win\_CC uses FP32, allowing im2win\_TC to process twice as many elements for the same data transfer volume. Second, tensor cores are optimized for matrix operations and can execute block-level computations in a single instruction, while CUDA cores perform scalar operations and handle one data element per instruction.

Im2win\_TC achieves a 1.4$\times$ improvement in overall TFLOPS compared to cuDNN\_TC and 6.4$\times$ compared to im2col\_cuBLAS\_TC. 
The underlying reasons for these performance gains will be examined in detail in the following subsection.

In terms of memory consumption, im2win\_TC uses on average about 53\% as much memory as cuDNN\_TC and approximately 35\% as much as im2col\_cuBLAS\_TC.
Im2win\_CC consumes less memory than cuDNN\_CC across all benchmarks except cv4, while im2win\_TC consumes less memory than cuDNN\_TC across all benchmarks except cv3 and cv4. 
The specific causes of this behavior will be analyzed in detail in the following subsection.
\subsection{Micro Benchmark with Each Convolution}
\begin{figure*}[!ht]
\centering
\includegraphics[width=\linewidth]{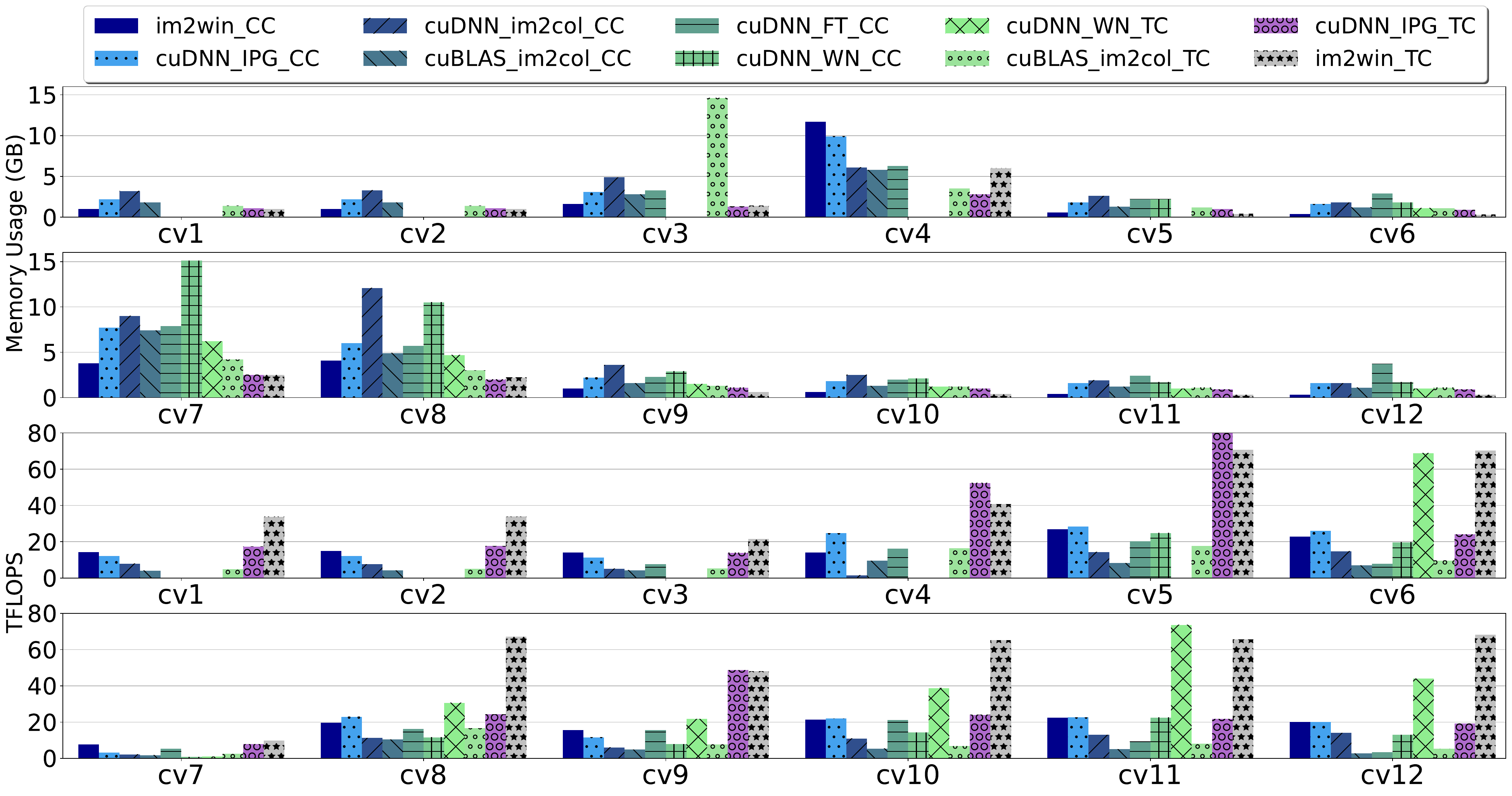}
\caption{ \small Performance results in TFLOPS and the memory usage of the im2win, PyTorch's im2col-based convolutions using cuBLAS, and cuDNN convolutions on GPU CUDA cores and tensor cores. Notes that some of these algorithms in cuDNN fail to run on some benchmarks due to certain constraints and TFLOPS of cuDNN\_IPG\_TC is 88.86 on cv5.}
\label{fig:7.3_Perfermance}
\end{figure*}

To explicitly compare our im2win implementations with each cuDNN convolution variant, we modify PyTorch’s convolution implementation to override the default adaptive algorithm selection and enforce a specific cuDNN algorithm for each benchmark. However, due to algorithmic constraints, certain algorithms may not be executable on specific layers. For details, please consult the cuDNN documentation.
We present the performance results in TFLOPS and the memory usage of the im2win, PyTorch's im2col-based convolutions using cuBLAS, and cuDNN convolutions on GPU CUDA and tensor cores in \cref{fig:7.3_Perfermance}.

On CUDA cores, the im2win algorithm achieves the highest TFLOPS on six convolutional benchmarks, while the cuDNN\_IPG implementation performs best on the remaining six benchmarks. For memory, im2win minimizes usage across all cases except cv4, which slightly exceeds cuBLAS\_im2col.

On Tensor Cores, im2win delivers peak TFLOPS on eight benchmarks; cuDNN\_IPG leads on cv3 and cv4, while cuDNN\_WN dominates cv11.

\subsubsection{Im2win vs cuBLAS\_im2col vs cuDNN\_im2col}
\label{batch strategies}
Overall, im2win\_CC delivers a 3.4$\times$ speedup over cuBLAS\_im2col\_CC and a 2.7$\times$ speedup over cuDNN\_im2col\_CC. Likewise, im2win\_TC achieves a 6.4$\times$ speedup compared to cuBLAS\_im2col\_TC.·

In terms of memory consumption, im2win\_CC achieves the lowest consumption across all benchmarks except cv4, on average about 62\% as much memory as cuBLAS\_im2col\_CC, about 41\% as much memory as cuDNN\_im2col\_CC. For im2win\_TC, it continues to maintain the smallest memory requirement on all remaining benchmarks except cv3, cv4 and cv8, on average about 55\% as much memory as cuBLAS\_im2col\_TC. 

The superior efficiency of im2win arises from its data transformation. Unlike the im2col approach, im2win flattens the convolution window, which reduces redundant data storage and enhances element reuse and spatial locality, as illustrated in \cref{fig:convolution}.

Recall there are three processing strategies when processing tensors in convolution: single-batch, mini-batch, or full-batch. The cuBLAS\_im2col implementation in PyTorch adopts a single-batch strategy, which allocates additional memory for an intermediate tensor corresponding to each batch. In contrast, im2win processes all batches concurrently, resulting in higher memory usage for certain cases such as cv4. Meanwhile, cuDNN\_im2col utilizes a mini-batch strategy, requiring more intermediate storage than the single-batch approach of cuBLAS\_im2col but less than full-batch processing.
\subsubsection{Im2win vs cuDNN\_IPG}
Compared to cuDNN’s implicit GEMM-based approach, im2win delivers comparable TFLOPS performance on CUDA cores. While on tensor cores, im2win achieves an overall 2$\times$ performance improvement over cuDNN. This performance gain primarily stems from the optimizations introduced in \cref{subsection:Overall Performance on GPU}, namely Zig-Zag access and asynchronous data movement.

Since cuDNN adopts a mini-batch strategy whereas im2win employs a full-batch approach for tensor processing, im2win incurs higher memory consumption than cuDNN’s implicit GEMM-based method in certain benchmarks. Nevertheless, on CUDA cores, im2win consistently consumes less memory than IPG across all benchmarks, averaging approximately 46\% of its memory usage. Excluding benchmarks cv3, cv4, and cv8, im2win\_TC further reduces memory consumption to an average of 60\% of cuDNN\_TC.

\subsubsection{Im2win vs cuDNN\_WN}
Compared to cuDNN\_WN\_CC, im2win\_CC achieves 2.5$\times$ higher overall TFLOPS, while im2win\_TC attains 2.8$\times$ overall TFLOPS of cuDNN\_WN\_TC. In terms of memory usage, 
im2win demonstrates substantially lower requirements; im2win\_CC consumes, on average, 27\% of the memory used by cuDNN\_WN\_CC, and im2win\_TC averages 37\% of the memory usage of cuDNN\_WN\_TC.
The Winograd algorithm significantly reduces the number of multiplications by introducing additional addition operations via linear transformation (see \cref{sec:related_work} for details). While this approach surpasses im2win in computationally small benchmarks, the associated transformations incur extra memory overhead, giving im2win a clear advantage in memory efficiency.
\subsubsection{Im2win vs cuDNN\_FT}\
Compared to the FFT-based approach im2win achieves 2.0$\times$ higher overall TFLOPS. In terms of memory consumptions, im2win requires on average 49\% of the memory required by cuDNN\_FT\_CC.

The FFT-based algorithm performs a transformation of the tensor from time to frequency domain (see \cref{sec:related_work} for details), which is the main computational overhead.
This overhead becomes more pronounced for larger tensor sizes, explaining why the FFT algorithm outperforms im2win on benchmark cv4. However, storing the intermediate tensor generated during the transformation introduces additional memory requirements, giving im2win lower overall memory overhead compared to the FFT-based approach.
\subsection{Ablation Study on Optimization Techniques}
\label{ablation study}
\begin{figure}[!ht]
	\centering
	\includegraphics[width=\columnwidth]{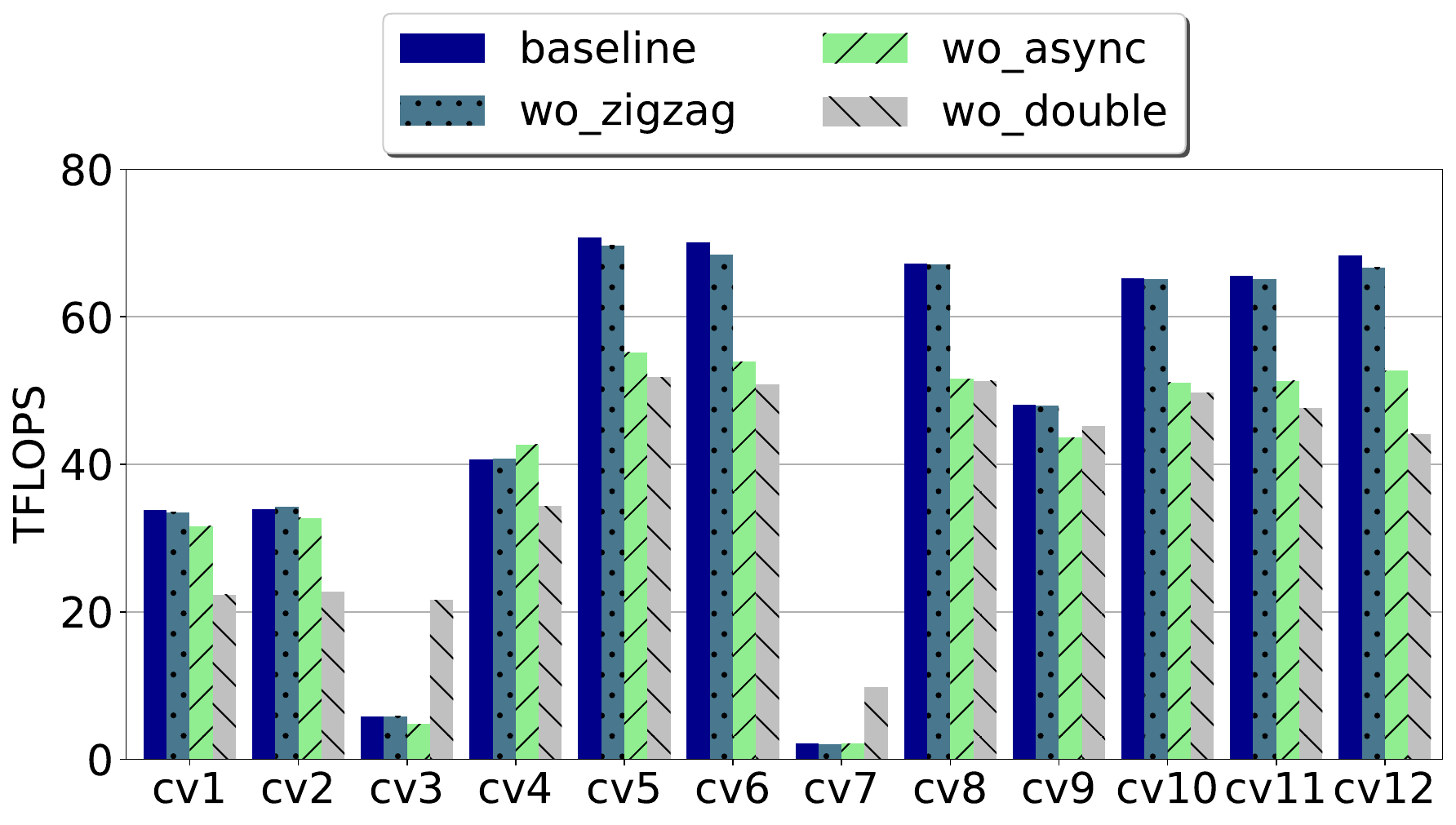}
	\caption{\small Performance results in TFLOPS of the ablation study of three optimization techniques on tensor cores.}
	\label{fig:ablation_study}
\end{figure}
To evaluate the individual impact of three optimization techniques, namely asynchronous data movement, double buffering and Zig-Zag access, we conducted an ablation study on the im2win convolution on tensor cores. 
The baseline includes all the proposed optimizations while three other variants disable one optimization at a time to evaluate its independent contribution.
\cref{fig:ablation_study} shows the performance results in TFLOPS across four optimization configurations. The results indicate that double buffering generally delivers the greatest performance gains, followed by asynchronous data movement, whereas Zig-Zag access yields the smallest improvement.

When assigning tasks to thread blocks, we typically chunk the convolution window, computing the elements corresponding to only one of the sub-blocks at a time. The computation proceeds in two phases: (i) loading input data from global memory into shared memory, and (ii) reading data from shared memory into registers for computation. Since the computation operation depends on the completion of the corresponding data load, we use the double buffering to implement the transfer of the next part of the data while computing the first part of the data, thus hiding the overhead of partial data transfer.

However, in benchmarks with small convolution windows, the performance is instead improved by removing the double buffering. This is mainly due to the fact that the double buffering itself introduces additional overheads, including latency from synchronization operations and additional share memory usage. When the convolution window is small, its inherent overheads outweigh its gains leading to a negative optimization of the double buffering.

In contrast, asynchronous data movement can also hide some of the data transfer latency, but its effectiveness is typically lower than that of double buffering. The Zig-Zag access pattern, designed primarily to reduce shared memory bank conflicts, delivers the most modest gains. Double buffering and asynchronous data movement have already reduced the probability of bank conflict to a certain extent through the data Pre-Fetching. Consequently, the marginal benefit of Zig-Zag access remains limited.

\section{Conclusion}

In this paper, we presented an optimized implementation of im2win convolution for modern GPUs, effectively addressing memory overhead and data locality challenges through a novel input tensor transformation. We extended the algorithm to support multi-precision computations, utilizing full-precision on CUDA cores and half-precision on tensor cores. %Guided by the Roofline model, 
Our kernel optimizations yield significant efficiency gains over standard approaches.
Our im2win implementation on CUDA cores shows significance performance improvement over other convolution variants in cuDNN and cuBLAS-based im2col.
%Experimental results demonstrate that on CUDA cores, im2win matches the performance of cuDNN and outperforms cuBLAS-based im2col by 3.6$\times$. On Tensor Cores, our approach surpasses cuDNN by 1.4$\times$ and cuBLAS by 6.4$\times$, while consuming only 53\% and 35\% of their respective memory footprints. Furthermore, the Tensor Core implementation achieves a 2.8$\times$ speedup over the CUDA-core version. 
Finally, our ablation study identifies double buffering as the most critical optimization factor, followed by asynchronous data movement, with zig-zag memory access providing supplementary performance benefits.

%The im2win convolution offers a promising solution by reducing memory usage and improving data locality through an innovative input tensor transformation. We extended it in this paper to support multi-precision computations on contemporary GPUs—using full-precision on CUDA cores and half-precision on tensor cores. 
%By incorporating proposed kernel optimizations guided by Roofline model, we demonstrate the possibility to attain significant performance and memory efficiency gains.

%On CUDA cores, im2win delivers up to 3.6$\times$ higher TFLOPS over im2col-based convolution with cuBLAS and matches cuDNN’s performance. On tensor cores, im2win delivers 1.4$\times$ higher TFLOPS than cuDNN and 6.4$\times$ higher than im2col-based convolution with cuBLAS, while requiring only 53\% and 35\% of their memory consumption respectively. Furthermore, the im2win implementation on tensor cores achieves 2.8$\times$ higher TFLOPS than the CUDA-core version.
%The ablation study shows that among the core optimization techniques applied to im2win convolution, double buffering has the greatest impact on performance, followed by asynchronous data movement, with zig-zag memory access contributing the least impact in comparison.

\section*{Acknowledgment}
This research was partially supported by the National Natural Science Foundation of China (Grant No. 42261070). This work is not related to Xu T. Liu's position at Amazon.

\bibliographystyle{IEEEtran}
\bibliography{bibliography}

\begin{thebibliography}{10}
\providecommand{\url}[1]{#1}
\csname url@samestyle\endcsname
\providecommand{\newblock}{\relax}
\providecommand{\bibinfo}[2]{#2}
\providecommand{\BIBentrySTDinterwordspacing}{\spaceskip=0pt\relax}
\providecommand{\BIBentryALTinterwordstretchfactor}{4}
\providecommand{\BIBentryALTinterwordspacing}{\spaceskip=\fontdimen2\font plus
\BIBentryALTinterwordstretchfactor\fontdimen3\font minus
  \fontdimen4\font\relax}
\providecommand{\BIBforeignlanguage}[2]{{%
\expandafter\ifx\csname l@#1\endcsname\relax
\typeout{** WARNING: IEEEtran.bst: No hyphenation pattern has been}%
\typeout{** loaded for the language `#1'. Using the pattern for}%
\typeout{** the default language instead.}%
\else
\language=\csname l@#1\endcsname
\fi
#2}}
\providecommand{\BIBdecl}{\relax}
\BIBdecl

\bibitem{shufflenet}
N.~Ma, X.~Zhang, H.-T. Zheng, and J.~Sun, ``Shufflenet v2: Practical guidelines
  for efficient cnn architecture design,'' in \emph{Proceedings of the European
  conference on computer vision (ECCV)}, 2018, pp. 116--131.

\bibitem{direct_conv_simd_ppopp_2023}
A.~d.~L. Santana, A.~Armejach, and M.~Casas, ``Efficient direct convolution
  using long simd instructions,'' in \emph{Proceedings of the 28th ACM SIGPLAN
  Symposium on Principles and Practice of Parallel Programming}, 2023, p.
  342–353.

\bibitem{mec}
M.~Cho and D.~Brand, ``Mec: memory-efficient convolution for deep neural
  network,'' in \emph{International Conference on Machine Learning}.\hskip 1em
  plus 0.5em minus 0.4em\relax PMLR, 2017, pp. 815--824.

\bibitem{ibxsmm}
A.~Heinecke, G.~Henry, M.~Hutchinson, and H.~Pabst, ``Libxsmm: accelerating
  small matrix multiplications by runtime code generation,'' in \emph{SC'16:
  Proceedings of the Int'l Conference for High Performance Computing,
  Networking, Storage and Analysis}.\hskip 1em plus 0.5em minus 0.4em\relax
  IEEE, 2016, pp. 981--991.

\bibitem{im2win_hpec_2022}
S.~Lu, J.~Chu, and X.~T. Liu, ``Im2win: Memory efficient convolution on {SIMD}
  architectures,'' in \emph{2022 IEEE High Performance Extreme Computing
  Conference (HPEC)}, 2022, pp. 1--7.

\bibitem{lu_im2win_2023}
S.~Lu, J.~Chu, L.~Guo, and X.~T. Liu, ``\BIBforeignlanguage{en}{Im2win: {An}
  {Efficient} {Convolution} {Paradigm} on {GPU}},'' in
  \emph{\BIBforeignlanguage{en}{Euro-{Par} 2023: {Parallel} {Processing}}},
  ser. Lecture {Notes} in {Computer} {Science}, Cham, 2023, pp. 592--607.

\bibitem{zhang2018high}
J.~Zhang, F.~Franchetti, and T.~M. Low, ``High performance zero-memory overhead
  direct convolutions,'' in \emph{Int'l Conference on Machine Learning}.\hskip
  1em plus 0.5em minus 0.4em\relax PMLR, 2018, pp. 5776--5785.

\bibitem{chellapilla_high_2006}
K.~Chellapilla, S.~Puri, and P.~Simard, ``High performance convolutional neural
  networks for document processing,'' in \emph{Tenth international workshop on
  frontiers in handwriting recognition}.\hskip 1em plus 0.5em minus 0.4em\relax
  Suvisoft, 2006.

\bibitem{p-im2col}
A.~V. Trusov, E.~E. Limonova, D.~P. Nikolaev, and V.~V. Arlazarov, ``p-im2col:
  Simple yet efficient convolution algorithm with flexibly controlled memory
  overhead,'' \emph{IEEE Access}, vol.~9, pp. 168\,162--168\,184, 2021.

\bibitem{evaluatcudnn_2019}
M.~Jordà, P.~Valero-Lara, and A.~J. Peña, ``Performance evaluation of {cuDNN}
  convolution algorithms on {NVIDIA} {Volta} {GPUs},'' \emph{IEEE Access},
  vol.~7, pp. 70\,461--70\,473, 2019.

\bibitem{qummar2025skepu}
S.~Qummar, A.~Ernstsson, C.~Kessler, and O.~Sysoev, ``Skepu-dnn: Algorithmic
  skeleton programming for deep learning on heterogeneous systems,'' in
  \emph{2025 IEEE International Parallel and Distributed Processing Symposium
  Workshops (IPDPSW)}.\hskip 1em plus 0.5em minus 0.4em\relax IEEE, 2025, pp.
  423--432.

\bibitem{im2win_2024}
X.~Fu, X.~Zhang, J.~Ma, P.~Zhao, S.~Lu, and X.~T. Liu, ``High performance
  im2win and direct convolutions using three tensor layouts on simd
  architectures,'' in \emph{2024 IEEE High Performance Extreme Computing
  Conference (HPEC)}, 2024, pp. 1--7.

\bibitem{dataLayout2}
B.~Li, Q.~Xue, G.~Yuan, S.~Li, X.~Ma, Y.~Wang, and X.~Tang, ``Optimizing data
  layout for training deep neural networks,'' in \emph{Companion Proceedings of
  the Web Conference 2022}, 2022, pp. 548--554.

\bibitem{dataLayout1}
C.~Li, Y.~Yang, M.~Feng, S.~Chakradhar, and H.~Zhou, ``Optimizing memory
  efficiency for deep convolutional neural networks on gpus,'' in \emph{SC'16:
  Proceedings of the International Conference for High Performance Computing,
  Networking, Storage and Analysis}.\hskip 1em plus 0.5em minus 0.4em\relax
  IEEE, 2016, pp. 633--644.

\bibitem{volkov2008benchmarking}
V.~Volkov and J.~W. Demmel, ``Benchmarking gpus to tune dense linear algebra,''
  in \emph{SC'08: Proceedings of the 2008 ACM/IEEE conference on
  Supercomputing}.\hskip 1em plus 0.5em minus 0.4em\relax IEEE, 2008, pp.
  1--11.

\bibitem{gpnpu}
Z.~Song, J.~Wang, T.~Li, L.~Jiang, J.~Ke, X.~Liang, and N.~Jing, ``Gpnpu:
  enabling efficient hardware-based direct convolution with multi-precision
  support in gpu tensor cores,'' in \emph{2020 57th ACM/IEEE Design Automation
  Conference (DAC)}.\hskip 1em plus 0.5em minus 0.4em\relax IEEE, 2020, pp.
  1--6.

\end{thebibliography}

\end{document}